\documentclass[prd,twocolumn,tightenlines,showpacs,nofootinbib]{revtex4-1} 

\usepackage{amsfonts}
\usepackage{dcolumn}
\usepackage{bm}
\usepackage{amsmath}
\usepackage{nicefrac}
\usepackage{amsthm}
\usepackage{amssymb}
\usepackage{graphicx}
\usepackage{color}

\newcommand{\appropto}{\mathrel{\vcenter{
  \offinterlineskip\halign{\hfil$##$\cr
    \propto\cr\noalign{\kern2pt}\sim\cr\noalign{\kern-2pt}}}}}

\begin{document}

\title{Can Dark Matter Induce Cosmological Evolution of the Fundamental Constants of Nature?}

\date{\today}
\author{Y.~V.~Stadnik} 
\affiliation{School of Physics, University of New South Wales, Sydney 2052, Australia}
\author{V.~V.~Flambaum} 
\affiliation{School of Physics, University of New South Wales, Sydney 2052, Australia}

\begin{abstract}
We demonstrate that massive fields, such as dark matter, can directly produce a cosmological evolution of the fundamental constants of Nature. 
We show that a scalar or pseudoscalar (axion-like) dark matter field $\phi$, which forms a coherently oscillating classical field and interacts with Standard Model particles via quadratic couplings in $\phi$, produces `slow' cosmological evolution and oscillating variations of the fundamental constants.
We derive limits on the quadratic interactions of $\phi$ with the photon, electron and light quarks from measurements of the primordial $^4$He abundance produced during Big Bang nucleosynthesis and recent atomic dysprosium spectroscopy measurements. These limits improve on existing constraints by up to 15 orders of magnitude. We also derive limits on the previously unconstrained linear and quadratic interactions of $\phi$ with the massive vector bosons from measurements of the primordial $^4$He abundance.

\end{abstract}


\maketitle 

\textbf{Introduction.} --- 
The idea that the fundamental constants of Nature might vary with time can be traced as far back as the large numbers hypothesis of Dirac, who hypothesised that the gravitational constant $G$ might be proportional to the reciprocal of the age of the Universe \cite{Dirac1937}. More contemporary dark energy-type theories, which predict the cosmological evolution of the fundamental constants, such as Brans-Dicke models, string dilaton models, chameleon models and Bekenstein models, assume that the underlying fields, which give rise to this evolution, are either massless or nearly massless, see e.g.~Refs.~\cite{Terazawa1981,Damour1994,Khoury2004,Bekenstein1982,Sola1987,Barrow2001,Pospelov2001,Barrow2008,Graham2013A,Martins2015,Marciano1984,Langacker2002,Calmet2002,Dent2003,Wetterich2003}. The evolution of the underlying dark energy-type field is determined by its couplings to matter, including dark matter, see e.g.~Ref.~\cite{Pospelov2001}. Another possible way of achieving a variation of the fundamental constants is via quantum effects induced by cosmological renormalisation group flow, see e.g.~Refs.~\cite{Shapiro2004,Shapiro2009,Fritzsch2012,Markkanen2014}.

In this letter, we demonstrate that a cosmological evolution of the fundamental constants can arise directly from a massive dark matter (DM) field, which is not necessarily unnaturally light. The possibility of exploring DM models in this particular context opens an exciting new avenue in the study of the cosmological evolution of the fundamental constants, since DM models are more amenable to theoretical and experimental investigation compared with their dark energy-type counterparts (see e.g.~Ref.~\cite{Bertone2010Book} and references therein). 

One of the leading candidates for DM is the axion, a pseudoscalar particle which was originally introduced in order to resolve the strong CP problem of Quantum Chromodynamics (QCD) \cite{Peccei1977,Peccei1977b} (see also \cite{Kim1979,Shifman1980,Zhitnitsky1980,Dine1981}). The axion forms a coherently oscillating classical field in the early Universe, which survives to the present day if the axion is sufficiently light and weakly interacting. These relic axions, which now reside predominantly in the observed galactic DM haloes, can be sought for through a variety of distinctive signatures, see e.g.~Refs.~\cite{Sikivie1983,ADMX2010,Graham2011,Sikivie2013LC,Stadnik2014axions,CASPER2014,Roberts2014prl,Roberts2014long,Sikivie2014atomic}. 
Apart from the QCD axion, one can also consider other spin-0 DM candidates, including scalar DM, the effects of which in astrophysical and cosmological contexts have been investigated extensively, see e.g.~Refs.~\cite{Gruzinov2000,Arvanitaki2010,Arvanitaki2011,Marsh2013,Rubakov2013,Schive2014A,Schive2014B,Porayko2014,Marsh2014A,Marsh2014B,Marsh2015}.

In the present letter, we consider a non-relativistic cold scalar or pseudoscalar (axion-like) DM field $\phi$, which is produced non-thermally (e.g.~through vacuum decay \cite{Preskill1983cosmo,Sikivie1983cosmo,Dine1983cosmo}) and forms a coherently oscillating classical field, $\phi = \phi_0 \cos(\omega t)$, that oscillates with frequency $\omega \simeq m_\phi c^2 / \hbar$, where $m_\phi$ is the mass of the DM particle, $c$ is the speed of light and $\hbar$ is the reduced Planck constant \cite{Footnote1}. In particular, although $\left<\phi\right> = 0$, $\left<\phi^2 \right> = \phi_0^2/2 \ne 0$ for such a field.  The DM model we consider in our present work satisfies constraints on both interaction strength and mass from existing experiments, including fifth-force searches and supernova energy loss bounds, and also satisfies gravitational requirements (including the formation of observed galactic DM haloes), since the non-gravitational interactions we consider are very weak.

\textbf{Theory.} --- 
The field $\phi$ can couple to the Standard Model (SM) fields via the following quadratic-in-$\phi$ interactions \cite{Footnote1b}:
\begin{align}
\label{quad_couplings_phi}
\mathcal{L}_{\textrm{int}} = &\mp \sum_{f}  \frac{ \phi^2 }{(\Lambda'_f)^2} m_f  \bar{f}f 
\pm \frac{ \phi^2 }{(\Lambda'_\gamma)^2} \frac{F_{\mu \nu} F^{\mu \nu}}{4} \notag \\
&\pm \sum_{V} \frac{\phi^2}{(\Lambda'_V)^2} \frac{M_V^2}{2} V_\nu V^\nu ,
\end{align}
where the first term represents the coupling of the scalar field to the SM fermion fields $f$, with $m_f$ the standard mass of the fermion and $\bar{f}=f^\dagger \gamma^0$, the second term represents the coupling of the scalar field to the electromagnetic field tensor $F$, and the third term represents the coupling of the scalar field to the SM massive vector bosons $V$, with $M_V$ the standard mass of the boson. 
Unlike the corresponding linear-in-$\phi$ interactions, here $\phi$ can represent either a scalar or pseudoscalar particle, since $\phi^2$ is of even parity for both.  

Comparing the terms in Eq.~(\ref{quad_couplings_phi}) with the relevant terms in the SM Lagrangian:
\begin{align}
\label{SM_Lagr+}
\mathcal{L}_{\textrm{SM}} = - \sum_{f}   m_f  \bar{f}f 
-  \frac{F_{\mu \nu} F^{\mu \nu}}{4} + \sum_{V}  \frac{M_V^2}{2} V_\nu V^\nu ,
\end{align}
we see that the SM particle masses are altered as follows:
\begin{align}
\label{delta-SM_masses+}
m_f \to ~&m_f \left[1 \pm \frac{\phi^2}{(\Lambda'_f)^2} \right] , 
~ M_V^2 \to M_V^2 \left[1 \pm \frac{\phi^2}{(\Lambda'_V)^2} \right] .
\end{align}
In order to see the effect of the quadratic coupling of $\phi$ to the electromagnetic field tensor, it is convenient to write the relevant terms in an alternate system of units:
\begin{align}
\label{alpha_varn_units+}
\mathcal{L} = -  \frac{F_{\mu \nu} F^{\mu \nu}}{4 e^2} \pm \frac{ \phi^2 }{(\Lambda'_\gamma)^2} \frac{F_{\mu \nu} F^{\mu \nu}}{4 e^2} ,
\end{align}
from which we deduce that the electromagnetic fine-structure constant $\alpha$ is altered as follows:
\begin{align}
\label{delta-FCs+}
\alpha \to \frac{\alpha}{1 \mp \phi^2 / (\Lambda'_\gamma)^2 } \simeq \alpha \left[1 \pm \frac{\phi^2}{(\Lambda'_\gamma)^2} \right] ,
\end{align}
provided that changes in $\phi$ are adiabatic.

Supernova energy loss arguments constrain the interaction parameters that appear in Eq.~(\ref{quad_couplings_phi}), for the scalar masses $m_\phi \lesssim T_{\textrm{core}}^{\textrm{SN}} \sim 30$ MeV. Consideration of the photon pair-annihilation channel $\gamma + \gamma \to \phi + \phi$ and nucleon bremmstrahlung channel $N + N \to N + N + \phi +\phi$
yield the limits: $\Lambda'_\gamma \gtrsim 3 \times 10^{3}~\textrm{GeV}$ and $\Lambda'_p \gtrsim 15 \times 10^{3}~\textrm{GeV}$, 
respectively \cite{Olive2008}. The limits on the interaction parameters in Eq.~(\ref{quad_couplings_phi}) from fifth-force searches are weaker, since for the quadratic couplings in Eq.~(\ref{quad_couplings_phi}), a fifth-force is produced in the leading order by the exchange of a pair of $\phi$-quanta between two fermions, which generates a less efficient $V(r) \simeq - m_f^2/ 64 \pi^3 (\Lambda'_f)^4 \cdot 1 / r^3$ attractive potential, instead of the usual Yukawa potential in the case of linear couplings. The strongest fifth-force limits are for the proton interaction parameter \cite{Olive2008}: $\Lambda'_p \gtrsim 2 \times 10^{3}~\textrm{GeV}$, for the scalar masses $m_\phi \lesssim 10^{-4}$ eV \cite{Adelberger2007+}. 

A lower limit on the mass of generic spin-0 particles, which saturate the observed cold DM content, comes from the requirement that the de Broglie wavelength of these particles not exceed the halo size of the smallest dwarf galaxies ($R \sim 1$ kpc): $m_\phi \gtrsim 10^{-22}$ eV (however, this limit is relaxed if spin-0 particles are not the dominant contributor to cold DM). This simple estimate is in fact in good agreement with more sophisticated astrophysical and cosmological considerations \cite{Marsh2013,Schive2014A,Schive2014B,Marsh2014A,Marsh2014B,Marsh2015}.

When $m_\phi \gg H(t)$, where $m_\phi$ is the mass of the DM particle and $H(t) \simeq 1/2t$ is the Hubble parameter as a function of time in the early Universe \cite{Rubakov_Book}, $\phi = \phi_0 \cos(m_\phi t)$ is an oscillating field and so $\phi^2$ contains both the oscillating term, $\phi_0^2 \cos(2 m_\phi t) / 2$, as well as the non-oscillating term, $\left<\phi^2\right> = \phi_0^2 / 2$. When $m_\phi \ll H(t)$, $\phi$ is a non-oscillating constant field due to the effects of Hubble friction. Thus, the temporal evolution of and spatial variations in $\left<\phi^2\right>$ produce `slow' space-time variations in the fundamental constants, which can be constrained from astrophysical phenomena, most notably Big Bang nucleosynthesis (BBN) and cosmic microwave background (CMB) measurements \cite{Footnote2}, while the oscillating component of $\phi^2$ produces oscillating variations in the fundamental constants, which can be sought for with high-precision laboratory measurements, such as atomic clock and laser interferometry experiments \cite{Stadnik2014laser,Stadnik2015laser}. 

The energy density of a non-relativistic oscillating DM field is given by $\rho \simeq  m_\phi^2 \left<\phi^2 \right>$ and evolves according to the relation:
\begin{equation}
\label{rho-temp}
\bar{\rho}_{\textrm{DM}} = 1.3 \times 10^{-6} \left[1 + z(t) \right]^3  \frac{\textrm{GeV}}{\textrm{cm}^3}  ,
\end{equation}
where $z(t)$ is the redshift parameter and the present mean DM energy density is determined from WMAP measurements \cite{PDG2012} (for relativistic DM, the mean DM energy density evolves as $\bar{\rho}_{\textrm{DM}} \propto [1+z(t)]^4$). 
The present-day cold DM energy density in our local galactic region is $\rho_{\textrm{CDM}}^{\textrm{local}} \approx 0.4~\textrm{GeV/cm}^3$ \cite{PDG2012}.
The energy density of a non-oscillating DM field is given by $\rho \simeq m_\phi^2 \left<\phi^2\right> / 2$ and, due to Hubble friction, is approximately constant while the field remains non-oscillating:
\begin{equation}
\label{rho-temp_OD}
\bar{\rho}_{\textrm{DM}} \simeq 1.3 \times 10^{-6} \left[1 + z(t_m) \right]^3  \frac{\textrm{GeV}}{\textrm{cm}^3}  ,
\end{equation}
where $z(t_m)$ is defined by $H(t_m) \approx m_\phi$. In both cases, the largest effect of variation of the fundamental constants induced by $\phi$, therefore, occurs during the earliest times of the Universe.

\begin{figure*}[h!]
\begin{center}
\includegraphics[width=8.5cm]{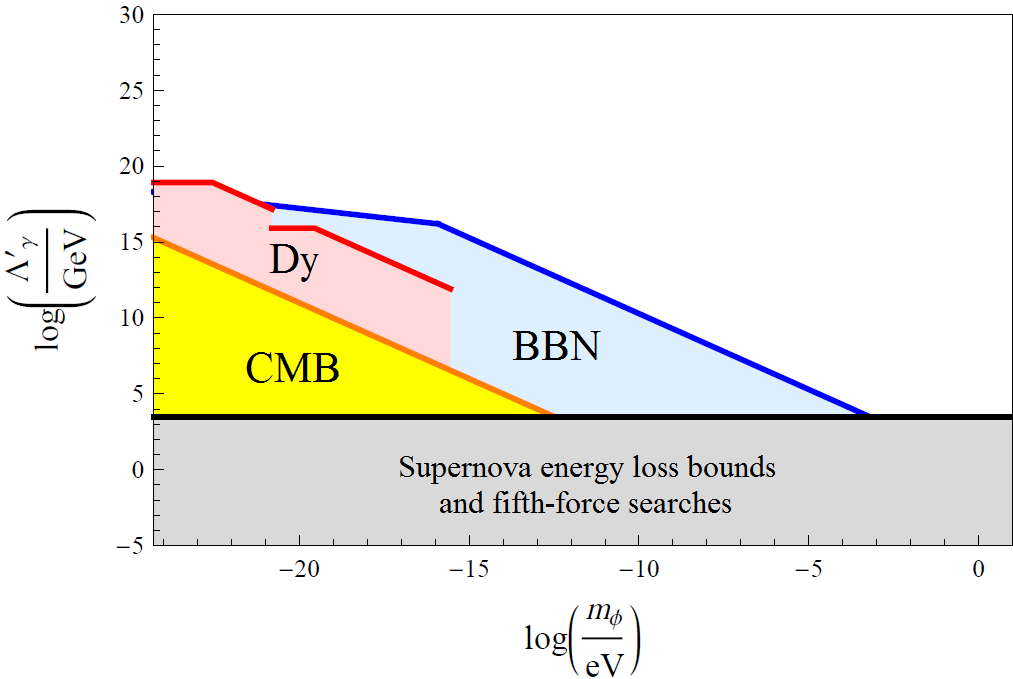}
\includegraphics[width=8.5cm]{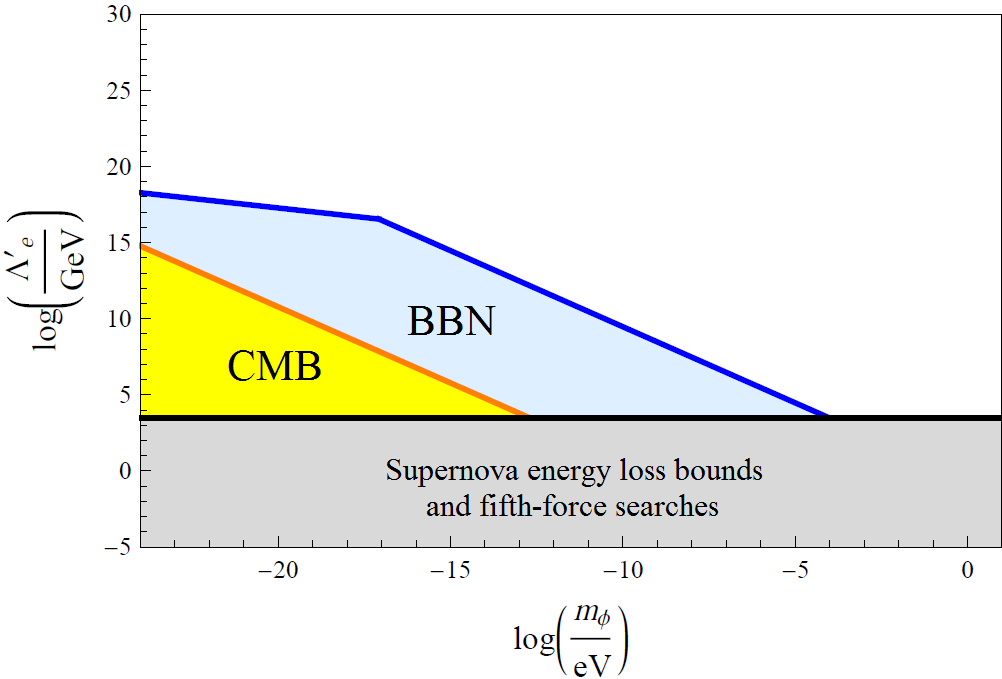}
\includegraphics[width=8.5cm]{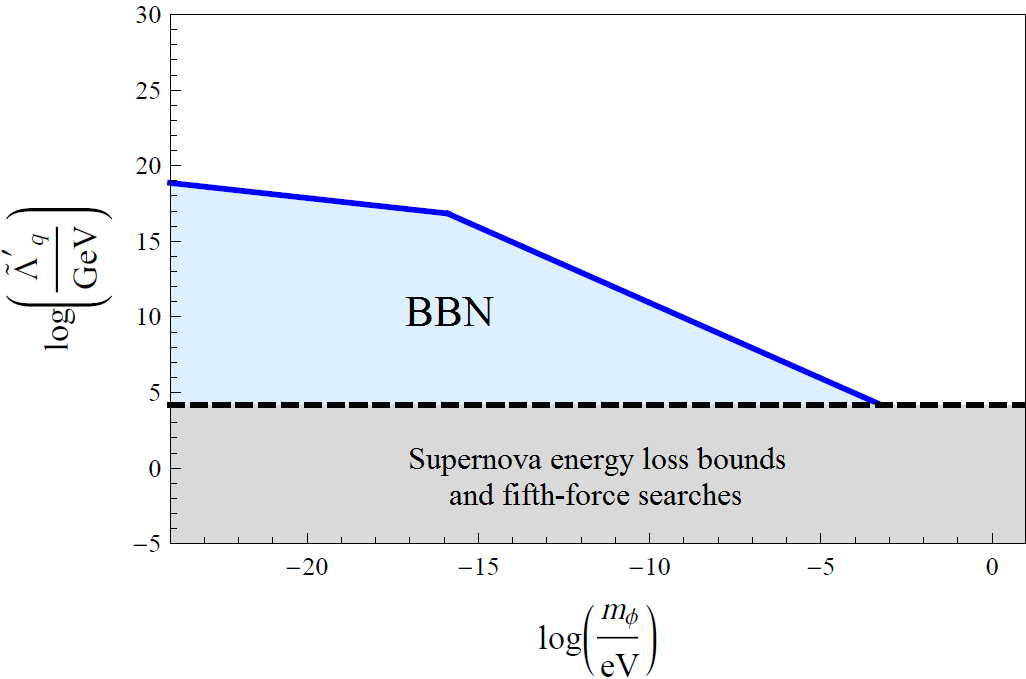}
\includegraphics[width=8.5cm]{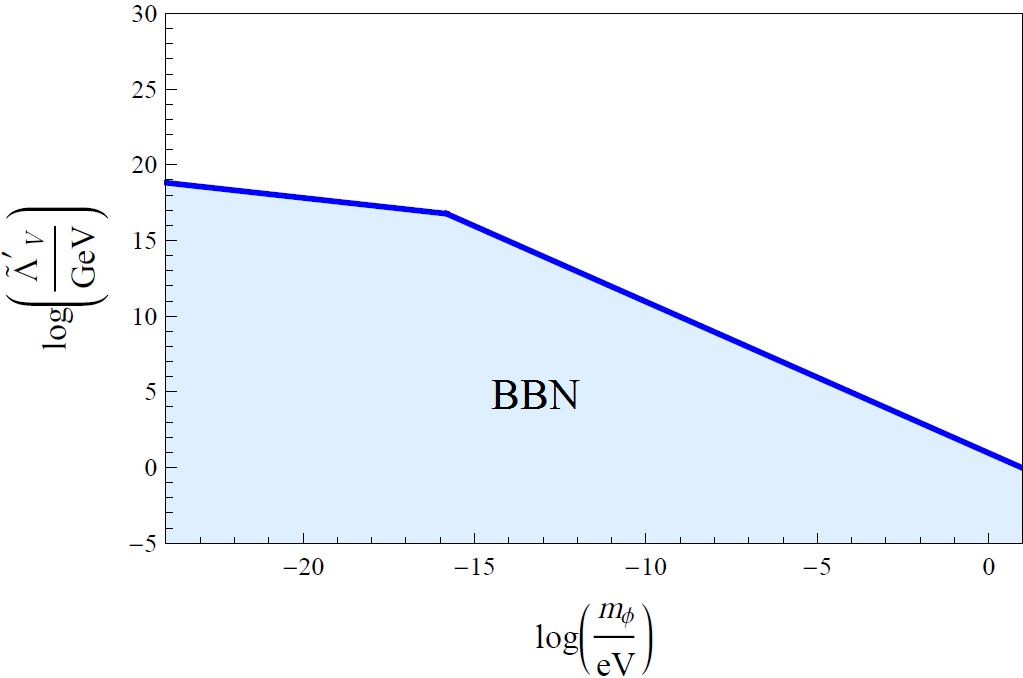}
\caption{(Color online) From top left to right:~Limits on the quadratic interactions of $\phi$ with the photon, electron, light quarks and massive vector bosons, as functions of the scalar particle mass $m_\phi$. 
Region below blue line corresponds to constraints derived in the present work from consideration of the primordial $^4$He abundance produced during BBN (the shape of the constraints on different sides of the mass $m_\phi \sim 10^{-16}$ eV arises from the different dependence of the scalar field amplitude $\phi_0$ on $m_\phi$ in these two limiting regions:~$\phi_0 \propto m_\phi^{-1} => \Lambda'_X \propto m_\phi^{-1}$ for $m_\phi \gg 10^{-16}$ eV, whereas $\phi_0 \propto m_\phi^{-1/4} => \Lambda'_X \propto m_\phi^{-1/4}$ for $m_\phi \ll 10^{-16}$ eV). 
Region below yelow line corresponds to constraints derived in the present work from consideration of CMB angular power spectrum measurements. 
Region below red line corresponds to constraints derived in the present work using recent atomic dysprosium spectroscopy data of Ref.~\cite{Budker2015}. 
Region below black line corresponds to existing constraints from consideration of supernova energy loss bounds and fifth-force experimental searches \cite{Olive2008}.
The light quark and massive vector boson interaction parameters are defined as $(\tilde{\Lambda}'_{q})^2 = |(\Lambda'_u)^2 (\Lambda'_d)^2 (m_d - m_u) / [(\Lambda'_u)^2 m_d - (\Lambda'_d)^2 m_u]|$ and $(\tilde{\Lambda}'_{V})^2 = |(\Lambda'_W)^2 (\Lambda'_Z)^2 / [(\Lambda'_Z)^2 - 1.4 (\Lambda'_W)^2]|$, assuming $\kappa'_d= \kappa'_u$ and $\kappa'_W= \kappa'_Z$.
} 
\label{fig:constraints_quad}
\end{center}
\end{figure*}

\textbf{CMB Constraints.} --- 
Variations in $\alpha$ and $m_e$ at the time of electron-proton recombination affect the ionisation fraction and Thomson scattering cross-section, $\sigma_{\textrm{Thomson}} = 8\pi / 3 \cdot \alpha^2 / m_e^2$, changing the mean-free-path length of photons at recombination and leaving distinct signatures in the CMB angular power spectrum. Recombination occurs over a relatively short period of time, $\Delta t_{\textrm{CMB}} / t_{\textrm{CMB}} \ll 1$ with $z(t_{\textrm{CMB}}) \approx 1100$, meaning that the energy density of DM is approximately constant during recombination. Analysis of WMAP measurements, which give the bounds $(\Delta \alpha / \alpha)_{\textrm{CMB}} \lesssim 0.01$ and $(\Delta m_e / m_e)_{\textrm{CMB}} \lesssim 0.04$ \cite{Landau2010}, hence immediately yield the following bounds on the quadratic interactions of the (oscillating) scalar field $\phi$ with the photon and electron:
\begin{equation}
\label{CMB_limits_quad}
\Lambda'_\gamma \gtrsim \frac{1~\textrm{eV}^2}{m_\phi}, ~ \Lambda'_e \gtrsim \frac{0.6~\textrm{eV}^2}{m_\phi},
\end{equation}
where we have made use of the fact that the mean DM energy density at the time of recombination is much greater than the present-day local cold DM energy density, and assumed that scalar DM saturates the present-day DM content. These constraints are presented in Fig.~\ref{fig:constraints_quad}.

\textbf{BBN Constraints.} --- 
Variations in the fundamental constants and particle masses in the early Universe alter the primordial abundances of the light elements. 
The most sensitive constraints come from consideration of the primordial abundance of $^4$He.
There are two distinct regions to consider --- the first is when the field $\phi$ is non-oscillating and approximately constant during BBN, which corresponds to the scalar mass range $m_\phi \ll 10^{-16}$~eV, while the second is when the field $\phi$ is oscillating for the entire duration of BBN, which corresponds to the scalar mass range $m_\phi \gg 10^{-16}$~eV.

We begin with the former case, when $\phi$ is constant. To leading order, changes in the primordial $^4$He abundance are given by:
\begin{equation}
\label{4He_reln_non-osc}
\frac{\Delta Y_p (^4\textrm{He})}{ Y_p (^4\textrm{He})} \approx \frac{\Delta (n/p)_{\textrm{weak}}}{ (n/p)_{\textrm{weak}}} - \Delta (\Gamma_n t_{\textrm{BBN}}) ,
\end{equation}
where $(n/p)_{\textrm{weak}} = e^{-Q_{np}/T_{\textrm{weak}}}$ is the neutron-to-proton ratio at the time of weak interaction freeze-out 
($Q_{np} = m_n - m_p = a\alpha \Lambda_{\textrm{QCD}} + (m_d - m_u)$ is the neutron-proton mass difference, with the present-day values $(a\alpha \Lambda_{\textrm{QCD}})_0 = -0.76$ MeV, where $\Lambda_{\textrm{QCD}} \approx 250$ MeV is the QCD scale and $a$ is a dimensionless constant, and $(m_d-m_u)_0 = 2.05$ MeV \cite{Leutwyler1982},
while $T_\textrm{weak} = b_1 M_W^{4/3} \sin^{4/3}(\theta_\textrm{W}) / (\alpha^{2/3} M_{\textrm{Planck}}^{1/3}) \approx 0.75$ MeV \cite{Rubakov_Book} is the weak interaction freeze-out temperature, $\theta_\textrm{W}$ is the Weinberg angle, $M_{\textrm{Planck}}$ is the Planck mass and $b_1$ is a dimensionless constant), $\Gamma_n$ is the neutron decay rate given by ($1/\Gamma_n = \tau_n \simeq 880$ s \cite{PDG2012}):
\begin{align}
\label{Gamma_n_formula}
\Gamma_n =  \frac{b_2 \alpha^2 m_e^5}{M_W^{4} \sin^{4}(\theta_\textrm{W})} \left[\frac{1}{15} (2x^4 - 9x^2 - 8) \sqrt{x^2 - 1} \right. \notag \\
+ \left. x \ln{(x + \sqrt{x^2 - 1})} \right] ,
\end{align}
with $x \equiv Q_{np}/m_e$ and $b_2$ a dimensionless constant, and $t_{\textrm{BBN}} \approx 180~\textrm{s}$ is the end-time of BBN.

The terms on the right-hand side of Eq.~(\ref{4He_reln_non-osc}) can be expressed in terms of variations of various fundamental constants:
\begin{align}
\label{n-p_varn}
&\frac{\Delta (n/p)_{\textrm{weak}}}{ (n/p)_{\textrm{weak}}} = 
-0.13 \frac{\Delta \alpha}{\alpha} - 2.7 \frac{\Delta (m_d - m_u)}{(m_d - m_u)} -5.7 \frac{\Delta M_W}{M_W}  \notag \\
&+8.0 \frac{\Delta M_Z}{M_Z}  +1.0 \frac{\Delta \Lambda_{\textrm{QCD}}}{\Lambda_{\textrm{QCD}}} -0.57 \frac{\Delta M_{\textrm{Planck}}}{M_{\textrm{Planck}}} ,
\end{align}
where the variations in all of the parameters are at the time of weak interaction freeze-out, and:
\begin{align}
\label{Gamma_n_varn_simple}
\frac{\Delta \Gamma_n}{ \Gamma_n } = -1.9 \frac{\Delta \alpha}{\alpha} +10 \frac{\Delta (m_d - m_u)}{(m_d - m_u)} - 1.5 \frac{\Delta m_e}{m_e}   \notag \\
 +10 \frac{\Delta M_W}{M_W}  -14 \frac{\Delta M_Z}{M_Z}  -3.9 \frac{\Delta \Lambda_{\textrm{QCD}}}{\Lambda_{\textrm{QCD}}} ,
\end{align}
where we have made use of the relation $\cos(\theta_\textrm{W}) = M_W/M_Z$, with $(M_W/M_Z)_0 = 0.882$ \cite{PDG2012}. In order to estimate the term $\Delta t_{\textrm{BBN}} / t_{\textrm{BBN}}$, we note that $t_{\textrm{BBN}} \propto M_{\textrm{Planck}} / T_{\textrm{BBN}}^2$ is determined from the condition that the rate of expansion and rate of strong interaction processes involved in BBN are equal. On dimensional grounds, we write $T_{\textrm{BBN}} = b_3 \Lambda_{\textrm{QCD}}^{1+y} / M_{\textrm{Planck}}^y \approx 60$ keV, where $b_3$ and $y$ are dimensionless constants. For $b_3 \sim 1$, we find $y \approx 0.2$, which gives:
\begin{equation}
\label{t_BBN_varn}
\frac{\Delta t_{\textrm{BBN}}}{t_{\textrm{BBN}}} \approx 1.4 \frac{\Delta M_{\textrm{Planck}}}{M_{\textrm{Planck}}} - 2.4 \frac{\Delta \Lambda_{\textrm{QCD}}}{\Lambda_{\textrm{QCD}}} .
\end{equation}

From the measured and predicted (within the SM) primordial $^{4}$He abundance, $Y_p^{\textrm{exp}} (^4 \textrm{He}) = 0.2477 \pm 0.0029$ \cite{Peimbert2007} and $Y_p^{\textrm{theor}} (^4 \textrm{He}) = 0.2486 \pm 0.0002$ \cite{Cyburt2008,Berengut2010}, we find the following constraints on the quadratic interactions of $\phi$ with the SM particles when $m_\phi \ll 10^{-16}$~eV, using Eqs.~(\ref{rho-temp_OD}), (\ref{4He_reln_non-osc}), (\ref{n-p_varn}), (\ref{Gamma_n_varn_simple}) and (\ref{t_BBN_varn}):
\begin{widetext}
\begin{align}
\label{constraints_non-osc_quad}
\frac{1}{m_\phi^2} \left(\frac{m_\phi}{3 \times 10^{-16} ~\textrm{eV}}\right)^{3/2}  &\left\{ \frac{0.25 \kappa'_\gamma}{(\Lambda'_\gamma)^2} + \frac{0.32 \kappa'_e}{(\Lambda'_e)^2} - \frac{4.9}{m_d-m_u} \left[\frac{\kappa'_d m_d}{(\Lambda'_d)^2} - \frac{\kappa'_u m_u}{(\Lambda'_u)^2} \right] - \frac{3.9 \kappa'_W}{(\Lambda'_W)^2}  + \frac{5.4 \kappa'_Z}{(\Lambda'_Z)^2} \right\} \notag \\
&\simeq  (-0.5 \pm 1.7) \times 10^{-20} ~\textrm{eV}^{-4} ,
\end{align}
where we have made use of the relation $[1+z(t_m)] / (1+z_{\textrm{weak}}) = \sqrt{t_{\textrm{weak}}/t_m}$ and the fact that the mean DM energy density during BBN is much greater than the present-day local cold DM energy density, and assumed that scalar DM saturates the present-day DM content. Here the $\kappa'_X = \pm 1$ correspond to the relevant signs in the Lagrangian (\ref{quad_couplings_phi}). These constraints are presented in Fig.~\ref{fig:constraints_quad}. Since the scalar field is non-oscillating during BBN when $m_\phi \ll 10^{-16}$~eV, we likewise also have the following constraints on the analogous linear interactions of $\phi$ with the SM particles:
\begin{align}
\label{constraints_non-osc_linear}
\frac{1}{m_\phi} \left(\frac{m_\phi}{3 \times 10^{-16} ~\textrm{eV}}\right)^{3/4}  &\left[ \frac{0.25 \kappa_\gamma}{\Lambda_\gamma} + \frac{0.32 \kappa_e}{\Lambda_e} - \frac{4.9}{m_d-m_u} \left(\frac{\kappa_d m_d}{\Lambda_d} - \frac{\kappa_u m_u}{\Lambda_u} \right) - \frac{3.9 \kappa_W}{\Lambda_W}  + \frac{5.4 \kappa_Z}{\Lambda_Z} \right] \notag \\
&\simeq  (-0.4 \pm 1.4) \times 10^{-11} ~\textrm{eV}^{-2} .
\end{align}
\end{widetext}

We now consider the case when $m_\phi \gg 10^{-16}$~eV, for which $\phi$ is oscillating during BBN. In this case, the only required modification in relation (\ref{4He_reln_non-osc}) is the replacement $\Delta (\Gamma_n t_{\textrm{BBN}}) \to \Delta \left[ {\int_{t_\textrm{weak}}^{t_{\textrm{BBN}}}} \Gamma_n (t) dt \right]$, where $t_\textrm{weak} \approx 1.1$ s. The dominant contribution to the variation of this integral comes from the variation of the integrand, which has the time dependence $\Delta \Gamma_n (t) \propto t^{-3/2}$, and from the variation of $t_{\textrm{weak}}$. 
Changes in the primordial $^4$He abundance due to changes in the neutron lifetime are thus suppressed by the small factor $2t_\textrm{weak}/t_\textrm{BBN} \approx 10^{-2}$ when $\phi$ is an oscillating field, compared with the case when $\phi$ is non-oscillating. From the measured and predicted (within the SM) primordial $^{4}$He abundance, we hence find the following constraints on the quadratic interactions of $\phi$ with the SM particles when $m_\phi \gg 10^{-16}$~eV, using Eqs.~(\ref{rho-temp}), (\ref{n-p_varn}), (\ref{Gamma_n_varn_simple}) and the modified version of Eq.~(\ref{4He_reln_non-osc}):
\begin{widetext}
\begin{align}
\label{constraints_osc_quad}
\frac{1}{m_\phi^2} \left\{ \frac{-0.13 \kappa'_\gamma}{(\Lambda'_\gamma)^2} + \frac{3 \cdot 10^{-3} \kappa'_e}{(\Lambda'_e)^2} - \frac{2.7}{m_d-m_u} \left[\frac{\kappa'_d m_d}{(\Lambda'_d)^2} - \frac{\kappa'_u m_u}{(\Lambda'_u)^2} \right] - \frac{2.9 \kappa'_W}{(\Lambda'_W)^2}  + \frac{4.0 \kappa'_Z}{(\Lambda'_Z)^2} \right\} \simeq  (-1.1 \pm 3.5) \times 10^{-20} ~\textrm{eV}^{-4} ,
\end{align}
\end{widetext}
where we have made use of the fact that the mean DM energy density during BBN is much greater than the present-day local cold DM energy density, and assumed that scalar DM saturates the present-day DM content. These constraints are presented in Fig.~\ref{fig:constraints_quad}.

\textbf{Atomic Spectroscopy Constraints.} --- 
We can derive constraints on the quadratic interaction of a coherently oscillating scalar field $\phi$ with the photon through the oscillating shifts it induces in atomic transition frequencies that are sensitive to variations in $\alpha$. Using the recent atomic dysprosium spectroscopy data of Ref.~\cite{Budker2015}, which were used to place constraints on the analogous linear interaction of $\phi$ with the photon, we derive constraints on the quadratic interaction of $\phi$ with the photon, assuming that scalar DM saturates the present-day local cold DM content. These constraints are presented in Fig.~\ref{fig:constraints_quad}.

\textbf{Spatial Variations in $\mathbf{\left< \phi^2 \right>}$.} --- 
We can derive constraints on spatial variations in $\left< \phi^2 \right>$ from measurements of spatial variations in the abundance of primordial deuterium, as determined from quasar absorption spectra: $d(D) = (5.4 \pm 2.9) \times 10^{-3} /$ Glyr \cite{Berengut2011} (we use the notation $d(X)$ to denote the fractional spatial gradient in parameter $X$). Without resorting to a full, time-dependent, numerical calculation, we assume that spatial variations in the deuterium abundance are due to spatial variations in $\left< \phi^2 \right>$ at the time of weak interaction freeze-out (for other mechanisms of generating spatial fluctuations in the primordial light elemental abundances, see Refs.~\cite{Kamion2003,Engelen2010}):
\begin{align}
\label{n/p_gradient_weak}
d(n/p)_{\textrm{weak}} &\sim d(D)_{\textrm{quasar}} \left(\frac{1+z_{\textrm{weak}}}{1+z_{\textrm{quasar}}}\right) \notag \\ 
& \sim (0.009 \pm 0.005)  ~\textrm{lyr}^{-1} ,
\end{align}
where $z_{\textrm{weak}} \approx 3.2 \times 10^9$ and $z_{\textrm{quasar}} \sim 1$. Assuming that scalar DM saturates the present-day DM content, and using Eqs.~(\ref{rho-temp}) and (\ref{n-p_varn}), this gives for $m_\phi \gg 10^{-16}$~eV:
\begin{widetext}
\begin{align}
\label{D_grad_osc_quad}
\frac{1}{m_\phi^2} \left(\frac{\nabla \rho}{\bar{\rho}}\right)_{\textrm{weak}} 
\left\{ \frac{-0.13 \kappa'_\gamma}{(\Lambda'_\gamma)^2}  - \frac{2.7}{m_d-m_u} \left[\frac{\kappa'_d m_d}{(\Lambda'_d)^2} - \frac{\kappa'_u m_u}{(\Lambda'_u)^2} \right] - \frac{2.9 \kappa'_W}{(\Lambda'_W)^2}  + \frac{4.0 \kappa'_Z}{(\Lambda'_Z)^2} \right\} \sim  (2.6 \pm 1.4) \times 10^{-20} ~\textrm{eV}^{-4} ~ \textrm{lyr}^{-1} .
\end{align}
Likewise, using Eqs.~(\ref{rho-temp_OD}) and (\ref{n-p_varn}), we find for $m_\phi \ll 10^{-16}$~eV:
\begin{align}
\label{D_grad_non-osc_quad}
\frac{1}{m_\phi^2} \left(\frac{\nabla \rho}{\bar{\rho}}\right)_{\textrm{weak}} \left(\frac{m_\phi}{3 \times 10^{-16} ~\textrm{eV}}\right)^{3/2}
\left\{ \frac{-0.13 \kappa'_\gamma}{(\Lambda'_\gamma)^2}  - \frac{2.7}{m_d-m_u} \left[\frac{\kappa'_d m_d}{(\Lambda'_d)^2} - \frac{\kappa'_u m_u}{(\Lambda'_u)^2} \right] - \frac{2.9 \kappa'_W}{(\Lambda'_W)^2}  + \frac{4.0 \kappa'_Z}{(\Lambda'_Z)^2} \right\} \notag \\
\sim  (1.3 \pm 0.7) \times 10^{-20} ~\textrm{eV}^{-4} ~ \textrm{lyr}^{-1} ,
\end{align}
\begin{align}
\label{D_grad_non-osc_lin}
\frac{1}{m_\phi} \left(\frac{\nabla \sqrt{\rho}}{\sqrt{\bar{\rho}}}\right)_{\textrm{weak}} \left(\frac{m_\phi}{3 \times 10^{-16} ~\textrm{eV}}\right)^{3/4}
\left[ \frac{-0.13 \kappa_\gamma}{\Lambda_\gamma}  - \frac{2.7}{m_d-m_u} \left(\frac{\kappa_d m_d}{\Lambda_d} - \frac{\kappa_u m_u}{\Lambda_u} \right) - \frac{2.9 \kappa_W}{\Lambda_W}  + \frac{4.0 \kappa_Z}{\Lambda_Z} \right] \notag \\
\sim  (1.1 \pm 0.6) \times 10^{-11} ~\textrm{eV}^{-2} ~ \textrm{lyr}^{-1} ,
\end{align}
where we have made use of the relation $[1+z(t_m)] / (1+z_{\textrm{weak}}) = \sqrt{t_{\textrm{weak}}/t_m}$.
\end{widetext}

\textbf{Conclusions.} --- 
We have demonstrated that massive fields, such as DM, can directly produce a cosmological evolution of the fundamental constants. 
We have shown that a (pseudo)scalar DM field $\phi$, which forms a coherently oscillating classical field and interacts with SM particles via quadratic couplings in $\phi$, produces `slow' cosmological evolution and oscillating variations of the fundamental constants.
We have derived limits on the quadratic interactions of $\phi$ with the photon, electron and light quarks from measurements of the primordial $^4$He abundance produced during BBN and atomic dysprosium spectroscopy measurements. These limits improve on existing constraints by up to 15 orders of magnitude. 
We have also derived limits on the previously unconstrained linear and quadratic interactions of $\phi$ with the massive vector bosons from measurements of the primordial $^4$He abundance.
Future laboratory experiments with atomic clocks, highly-charged ions, molecules and nuclear clocks (see e.g.~Refs.~\cite{Flambaum2008B,Ong2014} for overviews of possible systems), as well as laser interferometers \cite{Stadnik2014laser,Stadnik2015laser} offer a route for exploring new as-yet-unconstrained regions of physical parameter space.

\textbf{Acknowledgments.} --- 
We would like to thank Julian C.~Berengut, Gleb F.~Gribakin, Joan Sola and Ken Van Tilburg for helpful discussions. This work was supported by the Australian Research Council.


\begin{thebibliography}{99}
\bibitem{Footnote1b} Regarding other possible couplings of $\phi$ to the SM fields, we note that the effects of $\phi$, which we consider in the present work, on the Higgs vacuum expectation value $v \approx 250$ GeV are negligible, since the energy density of DM even at the time of weak interaction freeze-out is far too small ($\rho_{\textrm{DM}}^{\textrm{weak}} \ll v^4 \approx 4 \times 10^9$ GeV$^4$). The gauge coupling constant of QCD, $\alpha_{\textrm{strong}} \sim 1/\log(Q/\Lambda_{\textrm{QCD}})$, where $Q$ is the momentum transfer and $\Lambda_{\textrm{QCD}} \approx 250$ MeV is the QCD scale, is not defined for the energy scales considered in the present work, so we do not consider the coupling of $\phi$ to the gluon fields.
\bibitem{Footnote2} According to our estimates, the limits on the relevant DM interaction parameters from measurements pertaining to quasar absorption spectra, the Oklo natural nuclear reactor phenomenon and meteorite dating are not competitive with those from BBN and CMB measurements.
\end{thebibliography}



\end{document}